\documentstyle[aps,epsf,multicol]{revtex}

\newcommand{\beq}{\begin{equation}}
\newcommand{\eeq}{\end{equation}}
\newcommand{\bea}{\begin{eqnarray}}
\newcommand{\eea}{\end{eqnarray}}
\newcommand{\llabel}[1]{\label{#1}}

\newcommand{\half}[2]{\frac{\raisebox{-0.5ex}{\footnotesize $#1$}}
   {\mbox{\raisebox{0.25ex}[-0.2cm]{\footnotesize $#2$}}}}
\newcommand{\overset}[2]{\begin{array}{c} \raisebox{-0.7ex}{$~#1$} \\
\raisebox{1.8ex}{$#2$} \end{array}}

\begin{document}
\draft 
\preprint{}

\title{Uniform semiclassical expansions 
        for the direct part of Franck-Condon transitions}
\author{Bruno H\"upper and Bruno Eckhardt}
\address{Fachbereich Physik,
 Philipps Universit\"at Marburg, Renthof 6, 35032 Marburg, Germany}

\date{\today}
\maketitle
\begin{abstract}
Semiclassical expansions for traces involving Greens functions have
two contributions, one from the periodic or recurrent orbits of the
classical system and one from the phase space volume, i.e. the
paths of infinitesimal length. Quantitative calculations require
the control of both terms. Here, we discuss the contribution from
paths of zero length with an emphasis on the application to
Franck-Condon transitions. The expansion in the energy
representation is asymptotic and a critical parameter is identified.
In the time domain, a series expansion of the logarithm of the
propagator gives very good results.
The expansions are illustrated for
transitions onto a linear potential and onto a harmonic
oscillator.
\end{abstract}


\begin{multicols}{2}

\section{Introduction}
Many quantum properties, including the density of states and
Franck-Condon transition matrix elements can be expressed as
a trace of the Greens function times some operator
\cite{Llorente:1992,Heller:1978b}.
Semiclassical expressions for such quantities naturally divide up into
two parts, one due to ``paths of zero length'' and
one due to the longer recurrent or periodic trajectories of the
associated classical system
\cite{Berry:1983,Gutzwiller:1990,Eckhardt:1991b,Eckhardt:1992}
The numerical and conceptual
difficulties associated with the periodic orbit part have
received considerable attention in the literature
(see the contributions to \cite{Casati:1991,Cvitanovic:1993}
and the review \cite{Gaspard:1995}).
Zeta functions have helped to overcome many of these
problems, at least in certain, well-behaved situations
\cite{Gaspard:1995,Eckhardt:1995,Huepper:1997,Gaspard:1994}.
More recently, higher order corrections to the dominant
semiclassical contributions, in particular in the neighborhood of
caustics
and bifurcations have been addressed
\cite{Gaspard:1994,Gaspard:1993,Kus:1993,Schomerus:1997,%
Main:1997,Almeida:1987,Boasman:1995}.
In applications to the photodissociation of molecules one has
an additional source of corrections connected to the fact that
the operator is a projection on the initial state and hence
singular in the semiclassical limit. A way to deal with this
has been proposed by Zobay and Alber \cite{Zobay:1994}.

In mainly direct reactions the largest
part of the cross section comes, however,
from the paths of zero length. This leading order term is also
known as the Thomas-Fermi contribution for the case of
smooth systems or the Weyl-term for billiards. It measures
the volume of the energy shell in units of Planck's constant,
raised to the power of the number of degrees of freedom present
\cite{Berry:1983,Gutzwiller:1990}.
It turns out that in many situations one is too far away from
this semiclassical limit (typically one needs a higher density
of states) and so has to go beyond this leading order term. In
billiards,
the approximations to the density of states regularly contain
the subdominant contributions from the surface and corner corrections,
and even in smooth systems the leading order term alone will not do.
The expansion in decreasing powers of energy or wave number, however,
can typically be asymptotic at best. Building on their previous
developments in the theory of asymptotic series, the behavior
of the expansion has been illustrated
by Berry and Howls \cite{Berry:1994}
for the case of billiards:
the series expansion indeed diverges and the rate of divergence
is determined by short real or imaginary orbits of the classical system.

In view of this it is perhaps not surprising that in semiclassical
calculations of photodissociation cross sections one also has to
go beyond the leading order terms and that one encounters the same
kind of divergences \cite{Heller:1978b,Lee:1979,Huepper:1997}.
For practical applications the problems then are how to estimate the
importance of the higher order terms (without calculating them, of
course)
and how to improve on the series expansion. We will show here
that comparison to a simpler problem, namely excitation onto a
linear potential, suggests a useful parameter. For the second problem
we analyze in some detail the behavior of three different
approximations
to the background term and identify the most useful one. For the
sake of simplicity in notation all our calculations will be
confined to one degree of freedom only. Generalizations to more degrees
of freedom are straightforward. The main ideas will be illustrated for
transitions onto a linear potential and onto a harmonic potential.
Applications to photodissociation of water will be given elsewhere
\cite{HE:1998}.

Various theoretical aspects of the classical and semiclassical
limit of Franck-Condon transitions have been discussed previously
in the literature.
Much of the history is reviewed in the paper by
Dowling et al. \cite{Dowling:1991}, where also an interesting
alternative phase space interpretation for Franck-Condon transitions
can be found. Applications to molecules can be found
in the books \cite{Schi93,Child:1991}.
Of particular relevance
for our discussion is a paper by Heller \cite{Heller:1978b}
which contains background information
as well as a discussion of the first few correction terms; we summarize
some of his results in section IIA. Approximations in the time domain
(which we take up in section 3) have been discussed e.g. in
\cite{Fujiwara:1982,Makri:1988a,Makri:1988b}.

The outline of the paper is as follows. In section 2 we discuss the
Wigner- and the Grammaticos and Voros-expansion in the energy
representation.
The behavior at large orders for a linear potential is analyzed in
section 3.
In section 4 we
study approximations in the time domain, which can then be
connected to the energy domain by a Fourier transform (perhaps
to be evaluated numerically). The quality of the approximations is
illustrated for the harmonic oscillator in section 5. Some conclusions
will be
drawn in section 5.

\section{Semiclassical Franck-Condon factors}
\subsection{The Wigner series}
We consider transitions from an initial state $\Psi_i$,
typically a Gaussian, to
a manifold of final states $\Psi_E$ at energy $E$. The system has one
degree of freedom and the classical Hamiltonian on the upper potential
energy
surface is given by
\beq
H(p,q) = \half{p^2}{2m} + V(q)\,.
\llabel{Hamil}
\eeq
Quantum operators will be denoted with a '$\hat{\ }$',
so that
e.g. the quantum Hamilton operator will be $\hat{H}$.
The Frank-Condon factors we want to calculate are the
squares of the transition elements
\cite{Heller:1978b,Schi93,Child:1991},
\begin{equation}
  \rho(E) = \left|\langle \Psi_i | \Psi_E \rangle \right|^2 \,.
\end{equation}
Using $| \Psi_E \rangle \langle \Psi_E | = \delta(E - \hat{H})$,
where $\hat{H}$ is the Hamilton operator for the final electronic state,
we can write the Franck-Condon factor as
\begin{eqnarray}
  \rho(E) & = & \langle \Psi_i |\delta(E - \hat{H})|\Psi_i \rangle  \\
     & = &  \int_{-\infty}^{+\infty} dx\,
     \langle \Psi_i | x \rangle \langle x |
      \delta(E - \hat{H})|\Psi_i \rangle \\
   & = & \mbox{tr\,} \delta(E - \hat{H}) \hat{\Pi}  \, ,
\end{eqnarray}
 which has the above mentioned form as a trace over a Greens function,
 \bea
 \delta(E-\hat{H}) &=& -\half{1}{\pi}{\rm Im}~\lim_{\varepsilon\to 0}
     \hat{G}(E+i\varepsilon) \\
 &=& \half{1}{\pi} {\rm Im}~ \lim_{\varepsilon\to 0}
 {1 \over E-\hat{H}+i\varepsilon}
 \,,~~~~~\varepsilon > 0
 \eea
 times the projector $\hat{\Pi} = |\Psi_i \rangle\langle \Psi_i | $
onto
 the initial state. Taking the Wigner transform of this expression
 one arrives at classical phase space traces over the
 Wigner transforms of the operators involved \cite{Hillery:1984},
\begin{equation} 
      \rho(E)   = {1 \over h} \int_{-\infty}^{+\infty} d{p}
      \int_{-\infty}^{+\infty} d{q} \,
        [\delta(E - \hat{H})]_{W}(p,q)\, [\hat{\Pi}]_{W}(p,q) \, . 
\llabel{rho_cl}
\end{equation}
where the Wigner transform of an operator $\hat{A}$ is given by
\cite{Hillery:1984,Wigner:1932}:
\begin{equation} 
  [\hat{A}]_{W}(p,q)  =  
   \int d{x} \,   e^{-i{xp}/\hbar} \,
   \left\langle \left.q + \frac{x}{2}
   \right| \hat{A} \left|q - \frac{x}{2}\right.\right\rangle   \,.
\llabel{Wigprop}
\end{equation}
It is interesting to note that under the trace the Wigner transform of
the
product of two operators maps into the product of the Wigner transforms.
Generally, for a product of operators one has
\begin{equation}
	 [\hat{A}\hat{B}]_W(p,q) =  [\hat{A}]_W(p,q) \exp\left\{ \half{i\hbar}{2}
    \raisebox{-0.5ex}{\Large $\Lambda$}\right\} [\hat{B}]_W(p,q) 
\llabel{Wigequ}
\end{equation}
with the differential operator
\begin{equation}
	{\raisebox{-0.2ex}{\Large $\Lambda$}} =
 	\overset{\leftharpoonup~~}{\nabla_p}
       \overset{\rightharpoonup~~}{\nabla_q}   -
       \overset{\leftharpoonup~~}{\nabla_q}
       \overset{\rightharpoonup~~}{\nabla_p}   
\end{equation}
where the differentiations act to the left or to the right as indicated.
If the Hamilton operator is simply replaced by the classical Hamilton
function, the leading order term for the density of states, the
microcanonical measure on the energy shell,
$[\delta(E-\hat{H})]_W \sim \delta(E-H)$, is obtained.
The fact that formation of
products and Wigner transforms does not
commute is the source of quantum 
corrections which we want to study.

A systematic procedure to calculating an expansion of the
density of states in powers of $\hbar$ was suggested
long ago by Wigner \cite{Wigner:1932} and applied to the Franck-Condon
problem by Heller \cite{Heller:1978b}. It uses the statistical operator
$\hat{P}=\exp(-\beta\hat{H})$ and its high temperature expansion
for $\beta\to 0$. The statistical operator satisfies
\begin{equation}
- \half{\partial}{\partial\beta} \hat{P}
= \hat{H} \hat{P} = \half{1}{2}
\left( \hat{H} \hat{P} + \hat{P} \hat{H}\right) \, .
\llabel{P_eq}
\end{equation}
The symmetrized version on the right hand side is particularly
well suited for application of the Wigner transformation
as it immediately shows that the quantum corrections come
in even powers of $\hbar$ only.
Using (\ref{Wigequ}) for the product of two operators
the equation for $[\hat{P}]_W$ becomes
\begin{equation}
- \half{\partial}{\partial\beta} [\hat{P}]_W = [\hat{H}]_W
\cos\left\{ \half{\hbar}{2}
  \raisebox{-0.5ex}{\Large $\Lambda$}\right\} [\hat{P}]_W \,.
\llabel{PW}\llabel{pweq}
\end{equation}
The first few terms can be calculated by substitution of the
ansatz
\beq
[\hat{P}]_W = e^{-\beta H} \left(1 + \hbar^2 c_2 + \hbar^4 c_4 +\ldots \right)
\,.
\eeq
This expansion contains only even powers of $\hbar$ because of the
symmetrization in (\ref{P_eq}). Each coefficient $c_n$ is itself a
polynomial in $\beta$. For later reference we quote
$c_2$ for the Hamiltonian (\ref{Hamil}) \cite{Heller:1978b,Wigner:1932},
\beq
c_2 = -\beta^2 f_2 + \beta^3 f_3
\eeq
with
\bea
f_2 &=& \half{1}{8m} V'' \\
f_3 &=& \half{1}{24m} V'^2 + \half{1}{24m^2}V'' p^2
\llabel{f23}
\eea
where the primes denote derivatives with respect to position.

The statistical operator is the Laplace transform of the density of
states,
so to get back the quantity needed in the phase space trace, we need to
do
an inverse Laplace transform (see \cite{Heller:1978b} for more details).
The leading exponential then maps into
the delta function on the classical energy shell,
$\delta(E-H)$, and the powers of $\beta$ map into derivatives of the
delta function with respect to energy. The resulting expansion thus
becomes
\bea
[\delta(E-\hat{H})]_W &=& \delta(E-H) \nonumber\\
&\ & - \hbar^2 f_2
\half{\partial^2}{\partial E^2}\delta(E-H) + \hbar^2 f_3
\half{\partial^3}{\partial E^3}\delta(E-H)
\nonumber\\
&\ &+ \hbar^4 \ldots \,.
\llabel{Wigner_expansion}
\eea
Heller \cite{Heller:1978b} noted that instead of expanding in the form
of an exponential times a power series in $\beta$, one could
alternatively
expand as the exponential of a power series. For the terms given above
this leads to an Airy function approximation,
\bea
[\delta(E-\hat{H})]_W &=& \exp
\left\{-(H-E) f_2/3f_3 - 2\hbar^2 f_2^3 / 27 f_3^2\right\} \nonumber\\
& & \cdot\,\alpha\, \mbox{Ai}(\alpha(H-E+\hbar^2 f_2^2 / f_3))
\llabel{Airy_Heller}
\eea
with
\beq
\alpha = (3\hbar^2 f_3)^{-1/3} \,.
\llabel{alpha_Heller}
\eeq
Up to and including terms of order $\hbar^2$ the delta-function
expansion
of the Airy function in (\ref{Airy_Heller}) is equivalent to the
Wigner expansion (\ref{Wigner_expansion}).
We will come
back to this Airy function approximation below.

\subsection{The Grammaticos and Voros expansion}
The Wigner method of the previous section is
 a bit tedious when it comes to calculating
higher oder terms. A very convenient algebraic method of
expansion was developed by Grammaticos and Voros
\cite{Grammaticos:1979}.
The Dirac measure $[\delta(E - \hat{H})]_W$ is expanded
around the identity operator $\hat{I}$ times the classical
Hamilton function, $H(p,q) \cdot {\hat{I}}$.
The resulting series contains powers of the deviations
${\hat{H}} - H{\hat{I}}$,
which have an explicit $\hbar$ dependence because of the Wigner
equivalent of the product of two operators, Eq. \ref{Wigequ}.
To obtain this expansion, start from the integral representation,
\begin{eqnarray}
 & & \delta(E - \hat{H}) \nonumber \\ & = & \frac{1}{2\pi\hbar}
        \int_{-\infty}^\infty e^{iEt/\hbar} e^{-i\hat{H}t/\hbar} dt
           \\
    & = &
    	\frac{1}{2\pi\hbar}\int_{-\infty}^\infty
    		e^{i(E - H)t/\hbar} e^{-i(\hat{H} -H{\hat{I}}))t/\hbar} dt
    		\\
    & = & \frac{1}{2\pi\hbar}\int_{-\infty}^\infty
    	e^{i(E - H)t/\hbar} \sum_{r = 0}^\infty
          \frac{(-it)^r}{r!\hbar^r} ({\hat{H}} - H{\hat{I}})^r dt \,.
 \end{eqnarray}
Next,  take the Wigner transform on both sides and interchange
summation and integration,
\begin{eqnarray} 
 & & [\delta(E - \hat{H})]_W \nonumber \\ & = & \frac{1}{2\pi\hbar}\sum_{r =
0}^\infty
\frac{(-1)^r}{r!}
         \int_{-\infty}^\infty dt \left(\frac{it}{\hbar}\right)^r e^{i(E - H)t/\hbar}
         \left[ ({\hat{H}} - H{\hat{I}})^r\right]_W \llabel{texp} \\
       & = & \sum_{r = 0}^\infty \frac{(-1)^r}{r!}
       	\delta^{(r)}(E - H(p,q))\,
       {\cal{G}}_r(p,q,\hbar) \,. \llabel{dir}
\end{eqnarray}
The derivatives of the delta function are with respect to
the energy (as in (\ref{Wigner_expansion})) and the universal
coefficients ${\cal G}_r$ are the Wigner transforms of powers
of the Hamiltonian,
\begin{equation}
{\cal{G}}_r(p,q,\hbar) = \left[ ({\hat{H}} -
H(p,q){\hat{I}})^r\right]_W\,.
\end{equation}
They are universal in that other functions
of the Hamiltonian can be obtained by integration over energy,
e.g.
\bea
[f(\hat{H})]_W &=& \int_{-\infty}^{+\infty} dE\, f(E)\,
\delta(E-\hat{H})\\
&=& \sum_{r=0}^\infty  \frac{1}{r!} f^{(r)}(H(p,q))\,
{\cal{G}}_r(p,q,\hbar)
\,,
\llabel{f(H)}
\eea
and will contain the same coefficients. We will use this transformation
for the statistical operator and the Greens function below.

In principle, the ${\cal{G}}_r$ could be calculated by straightforward
expansions of the powers, and subsequent Wigner-transformation. However,
Grammaticos and Voros \cite{Grammaticos:1979} give a more efficient
algebraic technique, using Greens functions.
They consider the transformation (\ref{f(H)}) with
$f(\hat{H}) = 1/(\hat{H}-z)$ 
and thus find for Greens function
\begin{equation}
	[\hat{G}]_W(p,q,\hbar) = \sum_{r = 0}^\infty
    (-1)^r \frac{{\cal{G}}_r(p,q,\hbar)}{(H(p,q) -
    z)^{r+1} } \,.
\llabel{exp}
\end{equation}
The ${\cal{G}}_r$ can be read off from an expansion of
$[\hat{G}]_W(p,q,\hbar)$ in inverse powers of $(H - z)^{-r-1}$.
The Greens function satisfies $({\hat{H}} - z){\hat{G}}(z) =
{\hat{I}}$.
Using the symmetrized version as in (\ref{P_eq}) and
(\ref{PW}), this becomes in the Wigner representation
\begin{eqnarray}
 [\hat{I}]_W = 1 &=& 
	[({\hat{H}} - z){\hat{G}}(z)]_W
\label{G}
        \\
         & = & [({\hat{H}} - z)]_W\,
   {\rm cos}\left(
   \frac{\hbar}{2}\raisebox{-0.5ex}{\Large $\Lambda$} \right)
 \, [\hat{G}]_W(z)  \nonumber \\
   & = & (H - z)[\hat{G}]_W + \sum_{n=1}^\infty
\hbar^{2n}{\cal{H}}_{2n}
    [\hat{G}]_W \,.
\llabel{Hn}
\end{eqnarray}
The expansion in powers of $\hbar$ in the last line defines
differential operators ${\cal{H}}_n(p,q,\nabla_p,\nabla_q)$.
For $n = 2$, $4$ and the Hamiltonian (\ref{Hamil}) they are
\begin{eqnarray}
        {\cal{H}}_2 & = & -\frac{1}{8}\left( \frac{1}{m}
    \frac{\partial^2}{\partial q^2} + \frac{\partial^2V(q)}{\partial
q^2}
    \frac{\partial^2}{\partial p^2}\right) \,,
    \\
  {\cal{H}}_4 & = & \frac{1}{384} \frac{\partial^4V(q)}{\partial q^4}
  \frac{\partial^4}{\partial p^4} \,.
\end{eqnarray}
The Wigner transform of Greens function can be expanded in
powers of $\hbar$ as well,
\begin{equation}
[\hat{G}]_W(p,q,z,\hbar) = \sum_{m = 0}^\infty \hbar^m G_m(p,q,z)\,,
\llabel{G_hbar}
\end{equation}
and comparison of like powers of $\hbar$ in (\ref{G})
then gives
recursion relations for the coefficients $G_m$,
\begin{eqnarray}
        G_0(p,q,z) & = & \frac{1}{H - z } \,, \\
       G_{2n}(p,q,z) & = & -\frac{1}{H - z}\left(\sum_{k = 1}^n
       {\cal{H}}_{2k} G_{2(n-k)} \right) \,.
\llabel{Gexp}
\end{eqnarray}
Odd powers in $\hbar$ vanish because of symmetrization (as in
(\ref{P_eq},\ref{PW})).

Putting everything together one notes that by this method the
Greens function becomes a double series, organized in powers
of $\hbar$ and in inverse powers of $(H-z)$. Expression
(\ref{G_hbar}) emphasizes the former and (\ref{exp}) the latter,
so that the ${\cal G}_r$ can be obtained from (\ref{Gexp}) by
collecting contributions with the same powers of $(H-z)^{-r-1}$.
The structure of the expansion is such that there are always finitely
many contributions to the coefficients ${\cal G}_r$ (for an
example see the expansion for the harmonic oscillator in
(\ref{delta_ho}) below). The first few coefficients are
\begin{equation}
  {\cal{G}}_0 = 1 \,,
\end{equation}
\begin{equation}
  {\cal{G}}_1 = 0   \,,
\end{equation}
\begin{equation}
  {\cal{G}}_2 = -  \frac{\hbar^2}{4m} \frac{\partial^2 V(q)}{\partial
q^2} \,,
\end{equation}
\begin{eqnarray}
 {\cal{G}}_3 & = & - \frac{\hbar^2}{4m} \left( \frac{\partial
V(q)}{\partial q}
  \right)^2 - \frac{\hbar^2p^2}{4m^2}\frac{\partial^2 V(q)}{\partial
q^2}
  + \frac{3\hbar^4}{64m^2} \frac{\partial^4 V(q)}{\partial q^4} \,.
 \end{eqnarray}

Given these expansions, we now investigate their large order
behavior, first for the linear potential already studied by
Heller \cite{Heller:1978b}.

\section{Linear repulsive potential}
\subsection{The quantum case}
The simplest model for a Franck-Condon transition onto a
dissociating potential describes transitions from a Gaussian initial
state
\beq
\langle x | \Psi_i \rangle = \frac{1}{{\pi}^{1/4} \sigma^{1/2}}
e^{-(x-x_0)^2/2\sigma^2} \, ,
\llabel{Gauss0}
\end{equation}
onto a linear potential with Hamilton operator
\beq
\hat{H} = - \half{\hbar^2}{2m} \Delta - a x \,.
\llabel{H_lin}
\eeq
This model can be solved analytically 
and can be used as an approximation for transitions in an arbitrary
potential if $a$ is taken to be the slope of the upper potential
energy surface
at the maximum of the Gaussian initial state (which here is centered
at $q_0=0$). Lengths $\tilde{q} = q/\lambda$ can conveniently be
measured in
units of
\beq
\lambda = (\hbar^2/2ma)^{1/3}
\eeq
and energy in units of $a\lambda$, i.e $\epsilon = E/a\lambda$.
The eigenfunctions for this potential are Airy functions,
and $\lambda$ sets the scale for the width of the oscillations near
the turning point; the other oscillations in the wave function
have shorter wave length.
The Franck-Condon matrix elements can also be calculated exactly,
\begin{eqnarray} \rho(\varepsilon) & = & \frac{1}{a\sqrt{\pi}\sigma}
\left|
   \int_{-\infty}^\infty
    e^{-\tilde{q}^2s^2/2}     {\rm Ai}(-(\tilde{q} + \varepsilon))   
    d\tilde{q} \right|^2 \,,\label{Condlin} \\
\label{crosslinear}
  & = & \frac{2\sigma\sqrt{\pi}}{a\lambda^2} {\rm Ai}^2 \left(
     \half{1}{4} s^{-4} - \varepsilon \right) e^{s^{-6}/6 - \varepsilon
s^{-2}} \,.
\end{eqnarray}
They depend on energy and on the ratio
\beq
 s = \lambda / \sigma
 \llabel{s_def}
\eeq
of the length scale of the continuum wavefunction to the width of the
initial state.
This parameter also contains $\hbar$ and thus indicates how
`semiclassical' one is. For small $s$, i.e.  small $\hbar$ or
large $\sigma$ and a broad initial state,
the transition integral will average over many of the oscillations
of the Airy function and one can expect the leading classical term
to be reasonably accurate. However, for large $s$ and narrow Gaussians,
the initial state will probe every fine detail of the Airy function
and the classical approximation will presumably not work well.
We will come back to this point in section \ref{recur}.
It is our aim now to show how these quantum expectations are reflected
in the semiclassical expansion. As a first step we need to
calculate the terms in the series (\ref{dir}).

\subsection{Asymptotic expansions for a linear potential}
For the Hamiltonian (\ref{H_lin}) with its linear potential,
equation (\ref{Hn}) contains
only a single differential operator,
\begin{eqnarray}
	{\cal{H}}_2 & = & -\frac{1}{8}(H - z)\raisebox{-0.2ex}{\Large
	$\Lambda$}^2 \nonumber \\
       & = & -\frac{1}{8m} \frac{\partial^2}{\partial q^2}\,,
\nonumber
       \end{eqnarray}
all higher operators vanish.
Then the recursion relations for the $G_{2n}$
reduce to
\begin{equation}
	G_{2n} = -G_0 {\cal{H}}_2 G_{2(n-1)} \,.
\end{equation}
Since
\beq
G_0 = {1\over H-z} = {1\over
\frac{p^2}{2m} - a q - z}
\eeq
is linear in $q$, the action of ${\cal{H}}_2$ on powers
of $G_0$ is
\begin{equation}
	{\cal{H}}_2 G_0^n = -\frac{1}{8m}a^2 n(n+1) G_0^{n+2}\,.
\end{equation}
Together with the pre-factor $-G_0$ the coefficients $G_{2n}$ become
\begin{equation} 
	G_{2n} = \left(\frac{a^2}{8m}\right)^n
     \frac{(3n)!}{3^n n!} G_0^{3n+1} \,.
\end{equation}
Comparing the expansion
\begin{equation}
	 [\hat{G}]_W = \sum_{n = 0}^\infty  \left(\frac{a^2}{8m}\right)^n
	\hbar^{2n}
     \frac{(3n)!}{3^n n!} \frac{1}{(H - z)^{3n+1}}
\end{equation}
to the $\hbar$ - expansion of $[\hat{G}]_W$ in Eq. (\ref{exp})
one can easily read off the coefficients ${\cal{G}}_r$.
The asymptotic series (\ref{dir}) for the Dirac phase space measure
then becomes
\begin{equation}  
   [\delta(E - \hat{H})]_W =  \sum_{n = 0}^\infty \frac{1}{n!}
 	\left(\frac{a^2\hbar^2}{24m}\right)^{n} \delta^{(3n)}(E - H(x,p))
\,.
\llabel{delta_lin}
\end{equation}

This may be compared to the exact form of
$[\delta(E - \hat{H})]_W$ computed from the exact propagator,
\begin{eqnarray}
    [\delta(E & - & \hat{H})]_W \nonumber \\ & = & {\rm Ai} \left(\frac{E -
\frac{1}{2m}p^2 +
   aq}{\left(\frac{a^2\hbar^2}
    {8m}\right)^{1/3}}  \right) \frac{1}
    {\left(\frac{a^2\hbar^2 } {8m}\right)^{1/3}} \nonumber \\
    & = & \frac{1}{2\pi} \int_{-\infty}^\infty dz~\exp\left\{i(E-H)z -
       i\frac{a^2\hbar^2}{24m}z^3\right\} \,,
\end{eqnarray}
where $z$ is a dimensionless  integration variable.
The sequence of delta functions and their derivatives
may be obtained \cite{Heller:1978b} by expanding the exponential with
the
$z^3$ term
in a power series and interchanging integration and summation,
\begin{eqnarray}
 [\delta(E & - & \hat{H})]_W \nonumber \\
  & = & \frac{1}{2\pi} \sum_{n=0}^\infty \frac{1}{n!}
    \left(\frac{a^2\hbar^2}{24m}\right)^n \int_{-\infty}^\infty dz
      (-iz^3)^n e^{i(E-H)z} \nonumber \\
      & = & \sum_{n=0}^\infty \frac{1}{n!}
      	\left(\frac{a^2\hbar^2}{24m}\right)^n
         \delta^{(3n)}(E-H) \,.
\llabel{rho_lin}
\end{eqnarray}
These formulae can now be used for the calculation of the
Franck-Condon factors.

\subsection{Asymptotic behavior}
The Wigner transform of the initial Gaussian is a Gaussian in
phase space,
\begin{equation}
\Pi_{\mbox{\small $W$}}(q,p) = 2 \exp\left\{ {-
\frac{(q-q_0)^2}{\sigma^2} -
\frac{ \sigma^2 p^2}{\hbar^2} } \right\} \, ,
 \llabel{GauWigner}
\end{equation}
which helps to limit the domain of integration in
(\ref{rho_cl}) especially in cases
 of dissociation processes where the classical phase space is
unbounded in certain directions.
The expansion for the density of states,
eq. (\ref{rho_lin}), then gives for the series expansion of
the Franck-Condon factor
\begin{eqnarray}
  \rho_{\rm asymp}(\varepsilon) & = &  \frac{2}{\pi a \lambda} 
  \sum_{n = 0}^\infty  
   \frac{s^{3n}}{12^n n!} I_n(\varepsilon,s)  \label{rho_asympt}
\end{eqnarray}
with the integrals
\begin{eqnarray}  
	I_n(\varepsilon,s) & = &\int_0^\infty e^{-(x^2  -\varepsilon)^2
   s^2 - x^2/s^2} H_{3n}((x^2 -\varepsilon) s) dx \,.
   \label{rholinint}
\end{eqnarray}
Just as the exact quantum result the expansion contains only a
single parameter, $s=\lambda/\sigma$. The behavior for different
values of $s$ at $\varepsilon=0$, i.e. for the
maximum of the packet, is indicated in Fig.~\ref{asymptlin}.
For
small $s$ the terms decay rapidly and up to high $n$, but
eventually they start to grow and to diverge rapidly: this is the behavior
expected for an asymptotic series. As $s$ increases, the turnover
to divergence comes for smaller $n$ and for $s$ near $1$
all higher order terms are larger than the first one.
In Table \ref{convlin} we list the index of the smallest term
as well as its size for several values of $s$.

To estimate the rate of divergence of the series, one can substitute
the generating function for the Hermite polynomials and express
the integrals (\ref{rholinint}) for $\varepsilon=0$ as
\bea
I_n(s) &=& \left.\frac{\partial^{3n} }{\partial t^{3n}}\left(
 \int_0^\infty dx e^{ -(x^2 - \varepsilon)^2s^2 - x^2/s^2 + 2t(x^2 
 \varepsilon)s - t^2 }\right)\right|_{t = 0}\,.
\eea
This integral cannot be solved exactly. In saddle point approximation for
small $s$
one finds just a single saddle (for some $\varepsilon$ there are three).
The results is
\beq
I_n(s) =  \left.\frac{\partial^{3n} }{\partial t^{3n}}\left(
  \frac{\sqrt{\pi}}{2} \frac{s}{\sqrt{1 - 2ts^3}} e^{-t^2}\right)\right|_{t = 0} \,.
\eeq
Expanding in powers of $t$, one finally arrives at
\beq
I_n(s) = \frac{\sqrt{\pi}}{2} (3n)! \sum_{2l+r = 3n} \frac{(-1)^l}{l!}
  \left(2s^3\right)^{r+1} \left(\begin{array}{c} -1/2 \\ r \end{array} \right)\,,
\eeq
clearly showing a leading order behavior for small $s$ of the form
\beq
I_n(s) \approx
  \left\{\begin{array}{ll}
   {\large (-1)^{3n/2}\frac{\sqrt{\pi}}{2} \frac{(3n)!}{(3n/2)!} s } & n 
   \mbox{~even}, \\ & \\
  {\large (-1)^{(3n+1)/2}\frac{\sqrt{\pi}}{2} \frac{(3n)!}{((3n - 1)/2)!} s^4}
    & n \mbox{~odd}. 
   \end{array} \right.
\eeq
Together with the other prefactors in (\ref{rho_asympt}) a
typical term in the sum for $n$ even is
\beq
      \frac{(-1)^{3n/2}}{\sqrt{\pi}a\lambda}  
      \frac{(3n)!}{12^n~n!(3n/2)!} s^{3n+1}\,,
\eeq
and similarly for $n$ odd. With the help of Stirlings formula, one
can estimate the size of this term to be of the order
$\lbrack\sqrt{6} (n/12\,\mbox{e})^{1/3} s\rbrack^{3n}$. The terms start to
grow once the square bracket becomes larger than one,
i.e. for $n>n_c \approx (2\mbox{e}/\sqrt{6})\, s^{-3}$. 
This strong $s$ dependence is in accord with the variations in
Fig.~\ref{asymptlin}.

The information gained from this analysis of the linear potential 
can be used in more general settings: for excitations
into some arbitrary potential $V(q)$, the parameter $a$ can be
estimated from the slope of the potential at the maximum $q_0$
of the initial state, $a=-V'(q_0)$. If the parameter $s$ thus formed
is small, the classical approximation will be good (and the corrections
small), whereas some uniform approximation has to be tried if
$s$ is large.

\section{Time-domain approximations}
\subsection{The formal theory}

Another approach to the calculation of the mean density of states
exploits the relationship between energy- and time-domain. Since
the mean parts are obtained by averaging over (large) energy intervals,
they correspond to short time properties. This is the way the background
is obtained in numerical calculations: the wave packet is propagated for
a time long enough so that the overlap with the initial wave packet is
minimal but short enough so that no major part has returned, yet.
This finite time window is then Fourier transformed to obtain the
background term.

This can be mimicked in the semiclassical approximation.
The Franck-Condon factors are related to Feynman's propagator
in phase space by
\begin{equation}
  \rho(E) = \frac{1}{\pi\hbar} {\rm Re}\int_0^\infty dt
         \int \frac{ dq dp}{h} e^{iEt/\hbar}\, [\hat{\Pi}]_W(p,q)\,
[\hat{K]}_W(p,q) \,,
\end{equation}
where $[\hat{K}]_W$ is the Wigner transform of the  propagator
\begin{equation} 
  [\hat{K}]_{W}(p,q) = 
   \int d{x}\,    e^{-i{xp}/\hbar} 
   \left\langle q + \frac{x}{2}
   \right| e^{-i\hat{H}t/\hbar} \left|q - \frac{x}{2} \right\rangle  
\,.
\llabel{WigK}
\end{equation}
In the standard semiclassical approximation the smooth part
of the Franck-Condon factor is
obtained by use of the short time propagator from
${q - x/2}$ to ${q+ x/2}$ for Eq. (\ref{WigK}). In the
simplest approximation (e.g. Berry and Mount 1972), 
$\hat{K}$ is replaced by a propagator for a free particle
in a constant potential $V(q)$,
\begin{equation} 
  K({q + x/2},{q - x/2},t) = \left( \frac{m}{2\pi i \hbar t}
  \right)^{1/2}
         e^{i\left(\frac{m}{2t}{x}^2 - V({q})t
         \right)/\hbar}  \,.
\llabel{K0}
\end{equation}
The dominant part of the Franck-Condon factor as a function of the
energy
will come from that region in energy where the turning point of the
classical motion on the upper surface is near to the maximum of the
initial state. The ultimate form of this is the reflection
principle, where the wave function on the upper surface is
replaced by a delta function so that the cross section is
obtained by `reflection` of the initial state on the
potential energy surface (see \cite{Heller:1978b,Schi93}
for a discussion of its origin and its limitations).
This implies that the straight line propagator on a constant potential
is a rather
poor approximation since it does not account for a turning
point nor the exponential damping when entering the forbidden
region (See Fig.~\ref{tunnel}).

To obtain a better approximation linearize the potential
around the center point $q$ of the propagator,
\begin{equation}
  V({q}')  =  V({q + (q'-q)})
       \approx   V({q}) + V'({q})\,({q'-q}) \,,
\end{equation}
and use the exact propagator for the Lagrangian
\begin{equation}
  {\cal{L}}(\dot q', q') = \sum_i\frac{m_i}{2}\dot{q}'^2 -
V'({q})\,{q'}
             - V({q}) + V'(q)\,{q} \,,
\end{equation}
viz.,
\begin{eqnarray}
  & K\left({q + x/2},{q - x/2},t\right) = & \nonumber\\
  & \left({m \over 2\pi i \hbar t}\right)^{1/2}
               \exp\left( \frac{i}{\hbar} \left[ \frac{m}{2t} {x}^2
             - V({q}) - \frac{V'(q)^2}{24 m}t^3
             \right]\right) & \,.
\end{eqnarray}
The integration over $x$ and time gives an Airy function phase space
density,
\begin{equation}
\sigma_0(E) = \int \frac{d{p}d{p}}{h^N} ~
    \tilde\alpha\,{\rm Ai} \left(\tilde\alpha\,(H(q,p) - E)  \right)
    [\hat{\Pi}]_W(q,p)
\llabel{Airy}
\end{equation}
with
\begin{equation}
 \tilde\alpha = \left(\frac{\hbar^2|V'(q)|^2 }
 {8m}\right)^{-1/3} \,.
 \llabel{alpha_time}
\end{equation}
Comparison with the Airy function approximation introduced by Heller,
(\ref{Airy_Heller}) and (\ref{alpha_Heller}), shows that the two
coincide
for a linear potential: then $V''=0$ as well as $f_2=0$ and the
two scales (\ref{alpha_Heller}) and (\ref{alpha_time}) agree.

The above approximative formula for the direct part of Franck-Condon 
factors is exact for a linear potential and
contains an interesting limit case,
the one of an almost flat potential.

\subsection{Recurrence resonances}
\label{recur}

If the potential vanishes, $a=0$ in (\ref{H_lin}), the Gaussian
simply spreads and the Franck-Condon factor becomes
\begin{equation} 
  \rho_{\rm free}(E) = \sqrt{\frac{2m\sigma^2}{\pi \hbar^2}}
    \frac{e^{-2mE \sigma^2/\hbar^2}}{\sqrt{E}}
    \, .
\llabel{free}
\end{equation}
When $a = 0$ is substituted in (\ref{Airy}), the scale $\tilde\alpha$ becomes
infinite and the limit is singular. Similarly, the parameter
$s$ goes to infinity, indicating that the transition proceeds
via the full quantum regime. Note that in the
series expansion (\ref{delta_lin}) the transition can be performed
easily,
the coefficients of the derivatives of the delta function vanish. This
shows that the effect to be discussed here is non-perturbative.

On the classical side, the limit is also singular in a certain sense.
Fix the energy and consider the classical trajectories starting
at some point $q_0$. For vanishing potential, there are
two straight line trajectories running off to positive
or negative infinity, depending on momenta. As soon as there is
a sloping potential, no matter how small $a$, the topology changes
(see Fig.~\ref{proplin}).
For negative $a$, the trajectories running off to plus infinity will
pick
up in speed but be otherwise unchanged. The ones going towards
negative infinity, however, run up to the potential and will 
have to return after some time. The position of this turning point, $q_c = - E/a$
if $E$ is the kinetic energy at $q_0$, moves further out the smaller
$a$ and also the time for the trajectory to return increases with
decreasing $a$. The consequences of this are that the
return of the classical trajectory introduces a long time scale,
which, upon Fourier transform to the energy domain, will
manifest itself in a modulation on small energy differences.

These modulations are observed in the exact Franck-Condon factor 
(\ref{crosslinear}) in the case of large parameter $s$, as then
the structure from the Ai$^2$-part dominates (See Fig. \ref{fclin}).
A WBK quantization allows to connect the oscillations with
recurrent orbits.

The positions of the 'resonances' in the 
exact Franck-Condon factor eq.(\ref{crosslinear})
are given by the zeros of the first derivative
of the Airy function which are approximately described by \cite{Abramowitz:1984}:
\begin{equation}
\varepsilon_n = \left(\half{3\pi}{2}\left(n +
\half{1}{4}\right)\right)^{2/3}  
    ~\half{1}{4} s^{-4}    \label{zeroslin}
\end{equation}

If we quantize semiclassically the action of the closed (recurrent)
orbit which starts in the 
center of the Gaussian, is reflected by the potential and
returns to the initial point with reversed momentum by
requiring that
\begin{equation}
  S(E) = \frac{2}{\hbar} \int_0^{-\varepsilon\lambda} \sqrt{2m(E + ax)}dx
     = 
     2 \pi(n + \half{1}{4})
\end{equation}
(note that there is just one turning point, hence the contribution
$1/4$ instead of $1/2$ from the Maslov indices)     
we obtain  resonance positions 
\begin{equation}
\varepsilon_n^{sc} =
\left(\half{3\pi}{2}\left(n + \half{1}{4}\right)\right)^{2/3} \,.
\end{equation}
The quantization assumes orbits starting and ending in
the center of the Gaussian
and hence becomes more accurate for a narrow Gaussian.

In the limit $s \to 0$ these oscillations
are suppressed: The initial Gaussian has a momentum distribution
$e^{-\tilde{p}^2 s^{-2}}$ which is more concentrated around zero for small $s$.
If we regard the initial wavepacket as a cloud of particles with this
momentum distribution, it will move quite uniformly for small $s$, whereas for
large $s$ the cloud will spread quickly. The fast moving parts
are reflected and interfere with the slowly moving parts.
As the particles obey $\varepsilon = \tilde{p}^2 - \tilde{q}$, 
their energy distribution in the center of the Gaussian, $\tilde{q} = 0$,
is $e^{-\varepsilon s^{-2}}$.
This energy distribution factor is responsible for the damping of the
oscillations in the full cross section eq.(\ref{crosslinear}).

If we take a look at the exact autocorrelation function,
\begin{equation}
C
(t) = \frac{1}{\sqrt{1 + its^2}} e^{-t^2/4s^2 - it^3/12}\,,
\end{equation}
we have in the limit of the reflection principle 
\begin{equation}
C
(t) \stackrel{{s\to 0}}{\longrightarrow}
   e^{-t^2/4s^2}  \,,
\end{equation}
which results in a Gaussian cross section.
In the other limit 
\begin{equation}
C
(t) \stackrel{{s\to \infty}}{\longrightarrow}
        \frac{1}{\sqrt{1 + its^2}} e^{-it^3/12}  \,,
\end{equation}
the first part resembles the autocorrelation function of a freely moving
spreading wavepacket, but the second oscillating term
accounts for the recurrent path contributions to the propagator.

It is therefore reasonable to term the oscillating structures `recurrence 
resonances'. Although they are not connected to periodic orbits which imply
infinitely many revivals they are associated with recurrent orbits, and
their autocorrelation function differs from that of a simply spreading 
wavepacket.

Already Condon \cite{Condon:1928} and then later
Hunt and Child \cite{Hunt:1978} and Tellinghuisen \cite{Tellinghuisen:1984}
have mentioned these oscillations to appear on very
shallow dissociative potentials and described them with the help of an 
opposite reflection principle, namely that the continuum wave function
is reflected in the cross section \cite{Hunt:1978}: 
For large $s$, we have in the Franck-Condon factor eq.(\ref{Condlin}):
the initial state centered at $\tilde{q} = 0$ which acts like a 
needle scanning the continuum state:
\begin{equation}
   \frac{e^{-\tilde{q}^2s^2/2}}{\sqrt{2\pi}\sigma} 
       \stackrel{{s\to \infty}}{\longrightarrow} \delta(\tilde{q})
\end{equation}
and the cross section becomes
\begin{equation}
\rho(\varepsilon) = \frac{2\sigma\sqrt{\pi}}{V'\lambda^2} {\rm Ai}^2 \left(
     - \varepsilon \right)
\end{equation}
This example nicely illustrates the importance of recurrent orbits
since they are responsible for ALL the structures which often would
be associated with periodic orbits.

\subsection{Higher order approximations in the time domain}

In both the energy and the time domain, the Airy functions
are intuitively appealing approximations based on the first few
correction terms, but it is not obvious how to systematically
improve on them. A most convenient
method uses the Wigner expansion of the time propagator,
obtained from (\ref{f(H)}) with the Grammaticos and Voros method
and the universal coefficients ${\cal G}_r$, compare eq.~(\ref{dir}).
The Wigner transform of the
propagator has the expansion
\begin{equation}
  \left[e^{-i\hat{H}t/\hbar}\right]_W =
  \sum_{r = 0}^\infty \frac{(-it)^r}{r!\hbar^r}
  e^{-iHt/\hbar} {\cal{G}}_r({q},{p},\hbar)  \,,
\llabel{taylortexp}
\end{equation}
with the same ${\cal{G}}_r$ as calculated before.
This means that the autocorrelation function of the wave function,
\begin{eqnarray}
  C(t) &=& \langle\Psi_i | \Psi_r\rangle = {\rm tr\, }
e^{-i\hat{H}t/\hbar}
  \hat{\Pi} \\
  &=&
   \frac{1}{h} \int d{p} d{q} \,
        \left[e^{-i\hat{H}t/\hbar}\right]_{W}(p,q)\,
  [\hat{\Pi}]_{W}(p,q) 
  \llabel{c(t)}
\end{eqnarray}
can be constructed
with the help of this series.
The ${p}$-integrations can always be done analytically and we are
left only with the coordinate integrations.
The cross section is then obtained
by a Fourier transform,
\begin{equation}
	\rho(E) = \frac{2}{h} {\rm Re}~\int_0^\infty e^{iEt/\hbar} C(t) \chi(t)
	\,,
\end{equation}
where $\chi(t)$ is a window function that confines the region of
integration to the short times needed for the direct part.
This window is moreover important as the terms in the
expansion (\ref{taylortexp}) may
grow without bound for large time, influencing the convergence of the
integrals. As soon as
terms beyond the linear potential approximation are included the autocorrelation
function
may show recurrences structures which have to be switched off by means of a window.

Eq.~\ref{c(t)} has the advantage over 
(\ref{Wigner_expansion}) that it avoids the energy derivatives
which impose the highest demands on the
numerical accuracy of the phase space integrations. The
Fourier transform to energy can be done effectively using
fast Fourier transforms. However,
if the series expansion (\ref{taylortexp}) is used directly,
the high powers for large times have a devastating effect
numerically and nothing is gained. A way out of this
dilemma is suggested by the manipulations that lead
to (\ref{rho_lin}). Perhaps it is possible to sum
the power series expression in $t$ in (\ref{taylortexp})
into some exponentiated form, the first term of which would
be an Airy function.
This may be achieved using
\begin{equation}
   \sum_r \frac{(-it)^r}{r!\hbar^r}{\cal{G}}_r
   = \exp\left\{ -i\sum_k {\cal{F}}_k
   t^k \hbar^{-k}\right\} \,.
\end{equation}
The expansion coefficients ${\cal{G}}_r$ are known from the
algebra in the energy representation and
the coefficients ${\cal{F}}_k$ can be obtained from the
Plemelj-Smithies recursion relations
\cite{Plemelj:1909,Smithies:1941,Simon:1978}
\begin{equation}
  {\cal{F}}_n = \frac{(-1)^n i^{n+1}}{n!} {\cal{G}}_n -
   \frac{1}{n} \sum_{k = 1}^{n-1} k{\cal{F}}_k\frac{(-1)^{n-k}
i^{n-k}}
   {(n-k)!}{\cal{G}}_{n-k} \,.
\llabel{exponent}
\end{equation}
Since for $t=0$ we must have $\left[e^{-i(\hat{H} -
H)t/\hbar}\right]_W = 1$,
the coefficient ${\cal{F}}_0 = 0$. The next few are
\begin{eqnarray}
   {\cal{F}}_1 & = & 0 \,,\\
   {\cal{F}}_2 & = & -\frac{i}{2}{\cal{G}}_2 \,,\\
   {\cal{F}}_3 & = & -\frac{1}{6}{\cal{G}}_3 \,,\\
   {\cal{F}}_4 & = & \frac{i}{4}\left(\frac{1}{6} {\cal{G}}_4 -
   {\cal{G}}_2^2\right) \,.
\end{eqnarray}
The above formulas have an interesting phase space interpretation:
According to Berry \cite{Berry:1989b} we have now taken into account the
fringes in phase space surrounding the energy shell.
The difference to Berry's fringes-formula is that he
omits the second term in $t^2$ on the assumption that the stability
of the short trajectory segments does not change or that it
changes slowly compared to the other terms linear or
with time. He assumes the term quadratic
in $t$, which is connected with the stability of the short trajectory
segment,
to be constant.

\section{Matrix elements between harmonic oscillator eigenstates}

To illustrate the quality of the various approximations
we have to go beyond the linear potential (since there the
Airy functions are exact) and turn to excitations into harmonic
potentials.
Contrary to expectations based on the
usual close relationship between classical and
quantum dynamics in harmonic oscillators, the semiclasscial
expressions for the Franck-Condon transitions
are not exact and thus a useful test of
our formulae.

As in the linear case, the initial state $|\Psi_i\rangle$ 
is a Gaussian (\ref{Gauss0}) of width
$\sigma$ and a Gaussian Wigner transform in phase space,
(\ref{GauWigner}). The Hamiltonian now is
\beq
H = - \frac{\hbar^2}{2m} \Delta + \frac{m\omega^2}{2} q^2 \,.
\eeq
The length scale \mbox{$\ell = \sqrt{\hbar/m\omega}$}
is characteristic of the wave length of the ground state, all higher
excited states oscillate on shorter scales. It thus plays the same
role as the $\lambda$ in the case of the linear potential, setting
the largest scale for quantum oscillations.
The Franck-Condon transitions depend on the energy and on the single
parameter
\begin{equation}
r = \ell/\sigma = \sqrt{\frac{\hbar}{m \omega \sigma^2}} \,,
\llabel{r_def}
\end{equation}
which measures the size of the quantum oscillations relative to
the width of the initial state.

The Wigner transform of the initial state has a maximum in phase
space near $x=x_0$ and $p=0$. The leading order classical phase
space
average then yields a cross section which increases up to
$E_{max} \approx m \omega^2 x_0^2/2$ and decreases for higher energy.
Since the quantum spectrum is discrete, the quantum transition strengths
are discrete as well. A direct comparison between the leading order
classical Franck-Condon spectrum 
\beq
   \rho_0(E) = \int \frac{dqdp}{2\pi \hbar} \delta(E - H(p,q)) 
   \large[ \,,|\Psi_i\rangle\langle\Psi_i|\, \large]_W(p,q) \label{weylho}
\eeq
and the quantum data is shown in Fig.~\ref{HOobsmittel}.

The quantum calculations show that for large $\hbar$ and large
$r$ the quantum cross section is shifted towards higher energies.

The Grammaticos and Voros series (\ref{dir}) for the harmonic oscillator
can be calculated as before for the linear potential.
The asymptotic expansion of the phase space trace can be carried
out quite easily for the scaled Hamiltonian
$H = \hbar \omega
 \left(\frac{1}{2} \tilde{p}^2 + \frac{1}{2} \tilde{q}^2\right)$
where
$\tilde{q} = q/\ell$ and $\tilde{p} = p\ell/\hbar$.
 Again, there is only one operator in the series (\ref{Hn}),
 \beq
 {\cal{H}}_2 = \frac{1}{8} \frac{\partial^2}{\partial \tilde{q}^2}
 + \frac{1}{8} \frac{\partial^2}{\partial \tilde{p}^2}
 \eeq
 and the others vanish.
 From the recursion relations we then find up
 to order $O(\hbar^8)$ with the help of MAPLE:
\begin{eqnarray} 
&\large[& \delta(E - \hat{H}) \large]_W  = \delta(E - H) \nonumber \\
   &\ + & (\hbar\omega)^2\left( -\frac{1}{8}
    \delta''(E-H) + \frac{1}{12}E\delta'''(E-H)\right)  \nonumber\\
    &\ + &  (\hbar\omega)^4 \left(\frac{5}{384}\delta^{(4)}(E-H) 
         - \frac{3}{160}E~\delta^{(5)}(E-H)  \right. \nonumber\\ & & +
\left.
      \frac{160}{46080}E^2\delta^{(6)}(E-H) \right) \nonumber \\
     &\ + & (\hbar\omega)^6\left(-\frac{61}{46080}\delta^{(6)}(E-H) +
    \frac{479}{161280}E\delta^{(7)}(E-H) \right. \nonumber\\ & & -
\left.
     \frac{11648}{10321920}E^2\delta^{(8)}(E-H) -
    \frac{1120}{11612160}E^3\delta^{(9)}(E-H) \right) \nonumber \\
     &+& (\hbar\omega)^8 \left(\frac{1385}{10321920}\delta^{(8)}(E-H) -
    \frac{4757}{11612160}E\delta^{(9)}(E-H) \right. \nonumber \\ & &
\left. +
    \frac{1529}{6451200}E^2\delta^{(10)}(E-H) -
    \frac{17}{414720}E^3\delta^{(11)}(E-H) \right. \nonumber \\ & & +
\left.
     \frac{1}{497664}E^4\delta^{(12)}(E-H)\right)
\llabel{delta_ho}
\end{eqnarray}
which involves derivatives of the delta function up to 12th order.
Incidentally, this expansion shows rather clearly the structure of the
two different series for Greek's function, one in powers of
$\hbar$ (\ref{G_hbar}) and one in derivatives with respect to energy,
(\ref{dir}).
Fig.~\ref{hbarcorrosc} shows the
quantum corrections for the harmonic oscillator Franck-Condon factors
for different values of the expansion parameters.
Two trends can be recognized: For large $r$, divergence sets in very
early,
even from the first term on, and the smaller the energy (or the smaller
the
local gradient of the potential) the worse the
asymptotic expansion.
In both cases the leading order term gives the correct qualitative shape
of the Franck-Condon factor. For large energies, the higher order
corrections
are small and the series is close to the exact values. For lower
energies,
the higher order corrections increase and alternate in sign - the
typical
features of an asymptotic series.
The oscillations increase rapidly with $r$ so that the series becomes
practically useless without re summation.

The Airy correction (\ref{Airy}) has been computed
and gives remarkable results in the extreme quantum regime:
Fig.~\ref{corr4} shows the lowest order classical approximation to
the harmonic oscillator matrix elements and
the Airy function approximation. This confirms the importance
of turning point corrections and tunnel trajectories in the
Franck-Condon region.

\section{Final remarks}
We set out to calculate quantum corrections for Franck-Condon factors
in the classical phase space trace (\ref{rho_cl}). In 
a nutshell, there are three main results:
i) For the expansion based on (\ref{dir}) in the energy domain the
situation seems to be that either the corrections
are small to begin with (in which case one would be happy
with the leading order result) or they are large and
a resummation of the series is required as it is asymptotic at best.
ii) In the time domain the expansion in the exponent, (\ref{exponent}),
provides a more useful representation and contains non-perturbative
effects (the recurrence resonances).
iii) The quality of the approximation in all cases is controlled
by a single parameter, which may be estimated from the ratio
of a de Broglie wavelength to the width of the initial state
as in (\ref{s_def}) and (\ref{r_def}).

Clearly, the parameter controlling
the convergence can be put to immediate use and can help to
identify whether a classical approximation will suffice. An example
will be given in the photodissociation for water
\cite{Huepper:1997,HE:1998}.

The relationship between the series expansion in energy with its
derivatives of delta functions, (\ref{dir}) and the exponentiated
one in the time domain, (\ref{exponent}), is subtle and at
present not fully understood. The numerical observation
is that it is in very good agreement with the quantum results and
that no divergences seem to occur. The calculation leading to
(\ref{rho_lin}) for the expansion of the Airy function shows that
at least in this case all divergences are removed instantly:
the exponentiated series stops after the cubic in time. In
general circumstances, this will not be the case and the series
will continue. Then the question is whether the power series
in time, (\ref{taylortexp}), or its exponentiated version
is convergent or whether at some late time, perhaps related
to recurrent or periodic orbits, deviations between
quantum and semiclassical time evolution become noticeable.
For instance, in the case of the harmonic oscillator the
propagator has an exact recurrence after half a period which can
only be accounted for by a divergence of the series.
The success of the present calculation in the time domain would
then be related to the fact that the time evolution is followed
for a short time only. However, from a practical point of view,
that is all one is asking for in the calculation of the
direct part in Franck-Condon transitions, and
 we therefore propose to use
(\ref{c(t)}) with the exponentiated expansion (\ref{exponent})
and a fast Fourier transform to the energy domain.

The formulae presented allow for an accurate semiclassical
calculation of the background term and thus for the
largest contribution to the cross section in
predominantly direct reactions. The positions and widths of
resonances can be calculated from periodic orbit
expansions and zeta functions. Calculation of the
cross section however again requires
improved formulae which take into account the neighborhood of the
trajectories and the fringes in phase space are needed.
This would then allow to calculate all aspects of
cross sections semiclassically. Work along those lines is
in progress.

\section*{Acknowledgments}
We thank J.P. Keating for helpful comments on asymptotic series.
BE thanks the Newton Institute for Mathematical Sciences for
its hospitality during the writing of this article.
This work was supported by Deutsche Forschungsgemeinschaft.

\newcommand{\PR}[1]{{\em Phys.\ Rev.}\/ {\bf #1}}
\newcommand{\ACP}[1]{{\em Adv.\ Chem.\ Phys.}\/ {\bf #1}}
\newcommand{\AP}[1]{{\em Ann.\ Phys. (NY)}\/ {\bf #1}}
\newcommand{\CMP}[1]{{\em Commun.\ Math.\ Phys.}\/ {\bf #1}}
\newcommand{\JCP}[1]{{\em J.\ Chem.\ Phys.}\/ {\bf #1}}
\newcommand{\CPL}[1]{{\em Chem.\ Phys.\ Lett. }\/ {\bf #1}}
\newcommand{\JETP}[1]{{\em Sov.\ Phys.\ JETP}\/ {\bf #1}}
\newcommand{\JETPL}[1]{{\em JETP Lett.\ }\/ {\bf #1}}
\newcommand{\JMP}[1]{{\em J.\ Math.\ Phys.}\/ {\bf #1}}
\newcommand{\JMPA}[1]{{\em J.\ Math.\ Pure Appl.}\/ {\bf #1}}
\newcommand{\JPA}[1]{{\em J.\ Phys. A: Math. Gen. }\/ {\bf #1}}
\newcommand{\JPB}[1]{{\em J.\ Phys. B: At. Mol. Opt. }\/ {\bf  #1}}
\newcommand{\JPC}[1]{{\em J.\ Phys.\ Chem.}\/ {\bf #1}}
\newcommand{\PLA}[1]{{\em Phys.\ Lett.}\/ {\bf A #1}}
\newcommand{\PRL}[1]{{\em Phys.\ Rev.\ Lett.}\/ {\bf #1}}
\newcommand{\PRA}[1]{{\em Phys.\ Rev. A}\/ {\bf #1}}
\newcommand{\PRE}[1]{{\em Phys.\ Rev. E}\/ {\bf #1}}
\newcommand{\PRSL}[1]{{\em Proc.\ Roy.\ Soc.\ Lond. {\bf A}}\/{\bf #1}}
\newcommand{\PST}[1]{{\em Phys.\ Scripta }\/ {\bf T #1}}
\newcommand{\RMS}[1]{{\em Russ.\ Math.\ Surv.}\/ {\bf #1}}
\newcommand{\USSR}[1]{{\em Math.\ USSR.\ Sb.}\/ {\bf #1}}
\renewcommand{\baselinestretch} {1}

\narrowtext


\begin{figure}
\caption[]{Series expansion for Franck-Condon transitions onto
a linear potential for a Gaussian initial state.
Shown is the decadic logarithm of the relative error of the series
including the first $n$ terms
taken at the energy corresponding to the center of the Gaussian.
}
\llabel{asymptlin}
\end{figure}

\begin{figure}
\caption[]{Contributions to the Wigner propagator in position space.
The simplest approximation to the short time
  Wigner propagator uses the free one on a constant potential energy
$V({\bf Q})$. This is appropriate for high energies like $E_1$.
  Tunnel effects are important at lower energy $E_2$ and
  can be included by use of the propagator on the linearized 
potential.}
  \llabel{tunnel}
\end{figure}

\begin{figure}
\caption[]{Contributions to the Wigner propagator in phase space.
  Two paths contribute to the semiclassical propagator from 
	$Q-X/2$ to $Q-X/2$: the direct one 
  from phase space point 2 to 3
  and the path through the turning point from 1 to 3.}
  \llabel{proplin}
\end{figure}

\begin{figure}
\caption[]{
Exact Franck-Condon factors for the transition onto a linear
potential. For small $s$ the cross section eq. (\ref{Condlin}) has
a single maximum. For larger $s$ recurrence resonances develop.
In rescaled units they have a spacing given by eq. (\ref{zeroslin}). In
original units the oscillations become denser and denser.}
\llabel{fclin}
\end{figure}

\begin{figure}
\caption[]{
Comparison between the $n$-th harmonic oscillator eigenstate
matrix elements $\langle n | \Psi_i\rangle\langle
\Psi_i|n\rangle$ of the projector onto a Gaussian
$\Psi_i(q)$ and the corresponding
classical phase space average of the Wigner transform eq. (\ref{weylho}).
The center of the initial Gaussian remains the same for all figures.
The oscillator eigenvalues are $E_n = \hbar \omega(n + 1/2)$ with 
 $\hbar = 4.0,2.0,1.0,0.25$ and $\omega = 1.0$.}
\llabel{HOobsmittel}
\end{figure}

\begin{figure}
\caption[]{
  Quantum corrections to Frank-Condon factors for transitions to
  harmonic oscillator eigenstates.
  The different curves include all terms of eq. (\ref{delta_ho}) 
  up to the indicated order in
$\hbar$. For larger energies, the corrections improve the leading
order result. For low energy they develop oscillations and diverge
right away.}
  \llabel{hbarcorrosc}
\end{figure}

\begin{figure}
\caption[]{
 Comparison between the Weyl term, the uniform approximation and
 the exact results for transitions in harmonic oscillators.
 The exact result is shown by full circles, the Weyl term
 eq. (\ref{weylho}) as a solid line and the uniform approximation
 eq. (\ref{Airy}) as a dashed line.
}
 \llabel{corr4}
\end{figure}

\begin{table}
\caption{Asymptotics for the cross section of a Gaussian on a 
linear potential: We calculate the series up to its smallest term. 
The number of terms included
grows with decreasing expansion parameter $s$. ${\rm
log}_{10}\triangle\rho(\varepsilon = 0)$ is
the logarithm of the relative error of the series including the first
$n$ terms
taken at
the energy corresponding to the center of the Gaussian.
For large $s > 1$ the asymptotic series is practically of no use.}
 \llabel{convlin}

\begin{tabular}[t]{|c|c|c|}
\hline
 $~~ s ~~$ & $ ~~n ~~$ & ${\rm log}_{10}\triangle\rho(\varepsilon =
0)$  \\
 \hline
 1.8  &  0  &  0.03 \\
 1.6  &  1  &  -0.16 \\
 1.4  &  1  &  -0.59 \\
 1.2  &  1  &  -1.38 \\ 
 1.0  &  1  & -1.68 \\
 0.8  &  5 &  -2.50 \\
 0.63 &  21 &  -4.51 \\ \hline
 \end{tabular}
\end{table}

\end{multicols}
\end{document}